\documentclass[
twocolumn,
superscriptaddress,
floatfix,
showpacs,
amsmath,
amssymb,
aps,
prb,
]{revtex4-1}

\usepackage[]{graphicx} 
\usepackage{amsbsy} 
\usepackage{float}
\usepackage{color}
\usepackage[utf8]{inputenc}
\usepackage{dcolumn}
\usepackage{hyperref}

\begin{document}
\title{Electron-hole balanced dynamics in the type-II Weyl semimetal candidate WTe$_2$}

\author{M. Caputo}
\email{marco.caputo@u-psud.fr}
\affiliation{Laboratoire de Physique des Solides, CNRS, Universitè Paris-Sud, Universitè Paris-Saclay, 91405 Orsay Cedex, France}
\author{L. Khalil}
\affiliation{Laboratoire de Physique des Solides, CNRS, Universitè Paris-Sud, Universitè Paris-Saclay, 91405 Orsay Cedex, France}
\affiliation{UR1-CNRS/Synchrotron SOLEIL, Saint Aubin BP 48, Gif-sur-Yvette F-91192, France}
\author{E. Papalazarou}
\author{N. Nilforoushan}
\affiliation{Laboratoire de Physique des Solides, CNRS, Universitè Paris-Sud, Universitè Paris-Saclay, 91405 Orsay Cedex, France}
\author{L. Perfetti}
\affiliation{Laboratoire des Solides Irradiés, Ecole Polytechnique, CNRS, CEA, Universitè Paris-Saclay, 91128 Palaiseau Cedex, France}
\author{Q. D. Gibson}
\author{R. J. Cava}
\affiliation{Department of Chemistry, Princeton University, Princeton, New Jersey 08544, USA}%
\author{M. Marsi}
\affiliation{Laboratoire de Physique des Solides, CNRS, Universitè Paris-Sud, Universitè Paris-Saclay, 91405 Orsay Cedex, France}

\date{\today}

\begin{abstract}
We present a time- and angular-resolved photoemission (TR-ARPES) study of the transition-metal dichalcogenide WTe$_2$, a candidate type II Weyl semimetal exhibiting extremely large magnetoresistence.
Using femtosecond light pulses, we characterize the unoccupied states of the electron pockets above the Fermi level. We track the relaxation dynamics of photoexcited electrons along the unoccupied band structure and into a bulk hole pocket.
Following the ultrafast carrier relaxation, we report remarkably similar decay dynamics for electrons and holes.
Our results corroborate the hypothesis that carrier compensation is a key factor in the exceptional magnetotransport properties of WTe$_2$.
\end{abstract}


                              
\maketitle
\section{Introduction}
\begin{figure}[!h]
\includegraphics[width=\linewidth]{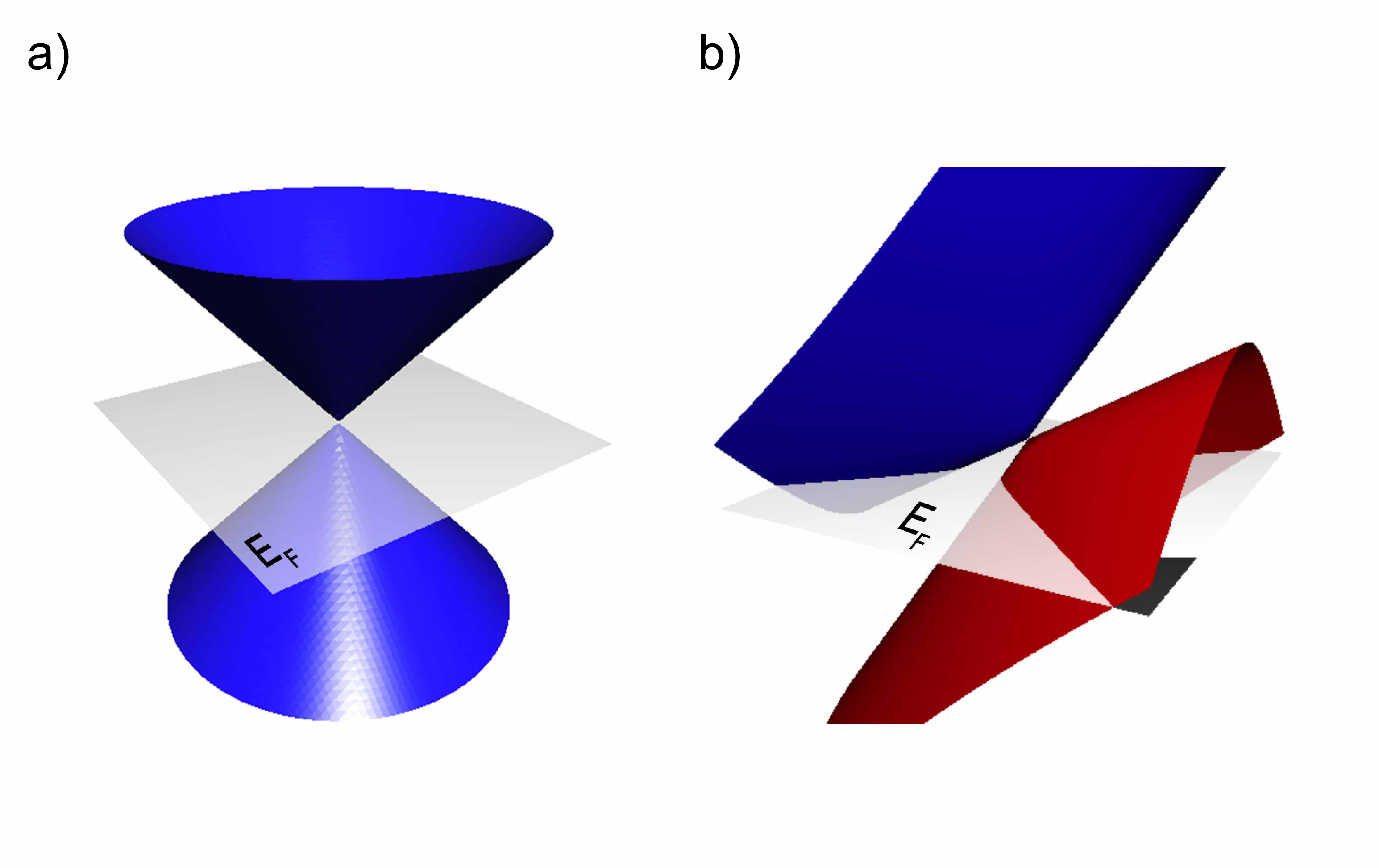} 
\caption{\label{Weyl}(Color online) (a) Schematic illustration of the type-I Weyl semimetal with point-like Fermi surface (the case of monopnoctide family), and (b) of type-II Weyl semimetal with tilted cone forming electron and hole pockets (the transition-metal dichalcogenide family)}
\end{figure}

Weyl semimetals establish a new class of quantum materials exhibiting topological protection \cite{lv2015,lv2015x,xu2015,liu2015}.
A point-touching in reciprocal space between valence and conduction band creates Weyl nodes associated to a topologically protected charge (Fig.~\ref{Weyl}(a)).
Their projection in the Fermi surface is point-like and the projection of nodes of opposite charge are connected with open surface states (Fermi Arcs)\cite{wan2011}.
Recently, a new class of Weyl semimetals was proposed, namely the type-II Weyl semimetals \cite{Soluyanov2015}, with Weyl points emerging from the linear crossing of electron and hole pockets.
The Weyl fermion, in this case, is characterized by a Weyl cone that is tilted over one side (Fig.~\ref{Weyl}b).
Violation of the Lorentz invariance in type-II Weyl semimetals is meant to lead to novel physical properties making these systems interesting to study \cite{Soluyanov2015}. 

Type-II Weyl fermions have been predicted to be found in some transition-metal dichalcogenides.
WTe$_2$ and MoTe$_2$ are among the first under investigation type-II Weyl semimetal candidates \cite{Soluyanov2015,Sun2015,Autes2016}.
High resolution ARPES measurements have successfully reveiled surface states connecting hole and electron pockets in WTe$_2$\cite{Pletikosic2014,Wang2016,Bruno2016,Wu2016,Wu2016pt,DiSante2017} and Mo$_x$W$_{1-x}$Te$_2$\cite{Chang2015,Deng2016,Huang2016,Tamai2016} along with their complete spin polarization\cite{Feng2016}.
Despite the intense efforts the topological nature of the surface states in these materials is still object of debate. TRAPRES experiments curried out on the unoccupied band structure of Mo-doped WTe$_2$ has unraveled arc-like surface states between the electron and hole pocket that could contain the topological states and, hence the Weyl points\cite{Belopolski2016a}.
Despite the excellent agreement between the calculated band structure and the experiment, no direct spectroscopic signature has been actually observed.
On the other hand, ultahigh resolution laser ARPES on thermally broadened WTe$_2$ showed a reach of surface states Fermi surface without, however, really conclusive results on its Weyl nature\cite{Liang2016,Wang2016}.
The Weyl points in this material reside above the Fermi level, thus making it difficult to be observed spectroscopically. 

The peculiar topology observed in various Dirac and Weyl materials has been proposed to be somehow related to their exceptional magnetotransport properties, like for instance the one found in Cd$_3$As$_2$ \cite{liang2015ultrahigh} and in NbP\cite{shekhar2015extremely}.
WTe$_2$ manifestically has been also known possessing an extremely large, nonsaturating magnetoresistance at low temperatures and high magnetic fields \cite{Ali2014}.
The high magnitude of its magnetoresistance has been attributed to an almost perfect compensation between electrons and holes and to the balanced arrangement of electron and hole pockets in the structure of the Fermi surface \cite{Pletikosic2014}.
To date, a possible relation between these phenemona and the topological nature of its surface states has not yet established.

In this paper, we report a TR-ARPES study of WTe$_2$ single crystals.
This technique makes it possible to explore the structure of its empty electronic states, and to observe the carrier relaxation dynamics after the system has been driven out-of-equilibrium by femtosecond optical pulses.
While our results cannot provide any conclusive evidence of Weyl nodes above the Fermi level, the relaxation dynamics of photoexcited states show a perfect balance between electrons and holes in the pockets at the proximity of the Fermi level.
This finding corroborates the hypothesis that electron-hole compensation is the key property to explain the extremely high magnetoresistance.

\begin{figure}[t]
\includegraphics[width=\linewidth]{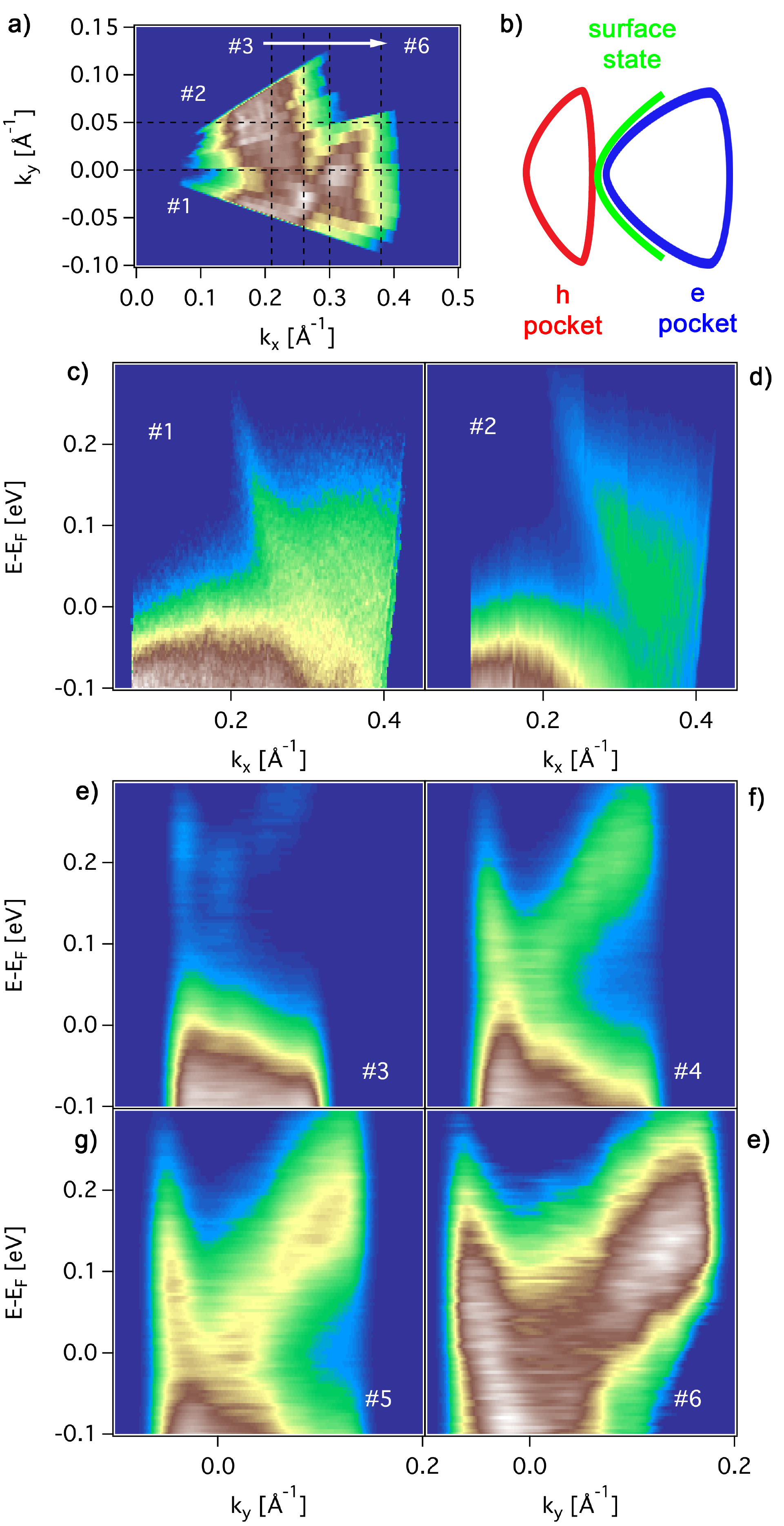}
\caption{\label{fermi} (Color online) (a) Fermi surface  of the WTe$_2$ along with its schematic illustration (b). Cuts along the high symmetry direction $\Gamma$X (c) and parallel to it (d), and cuts parallel to the high symmetry direction $\Gamma$Y (e)-(f) (lines \#1--\#6 in (b)).}
\end{figure}

\section{Methods}
High quality WTe$_2$ single crystals were synthesized by the flux method as described elsewhere \cite{Ali2014}.
The pump-probe photoemission experiments have performed at the FemtoARPES setup consisting of a commercial Ti:Sapphire laser delivering 35~fs pulses at 1.57~eV at a repetition rate of 250~kHz.
Part of the output laser power was used to generate probe 6.28 eV photons through cascade frequency mixing in BBO crystals \cite{Faure2012}.
The energy and time resolution were estimated to be of $\approx$50 meV and 150 fs, respectively.
The used pump fluence was 0.6$\pm$0.02 mJ/cm$^2$ and the photoemission spectra were referred to the Fermi level ($E_\text{F}$) of the system at thermodynamic equilibrium.
All samples were cleaved \textit{in~situ} at room temperature and base pressure of 1$\times$10$^{-10}$ mbar before being cooled down to 130 K. 

\begin{figure*}[t]
\includegraphics[width=\textwidth]{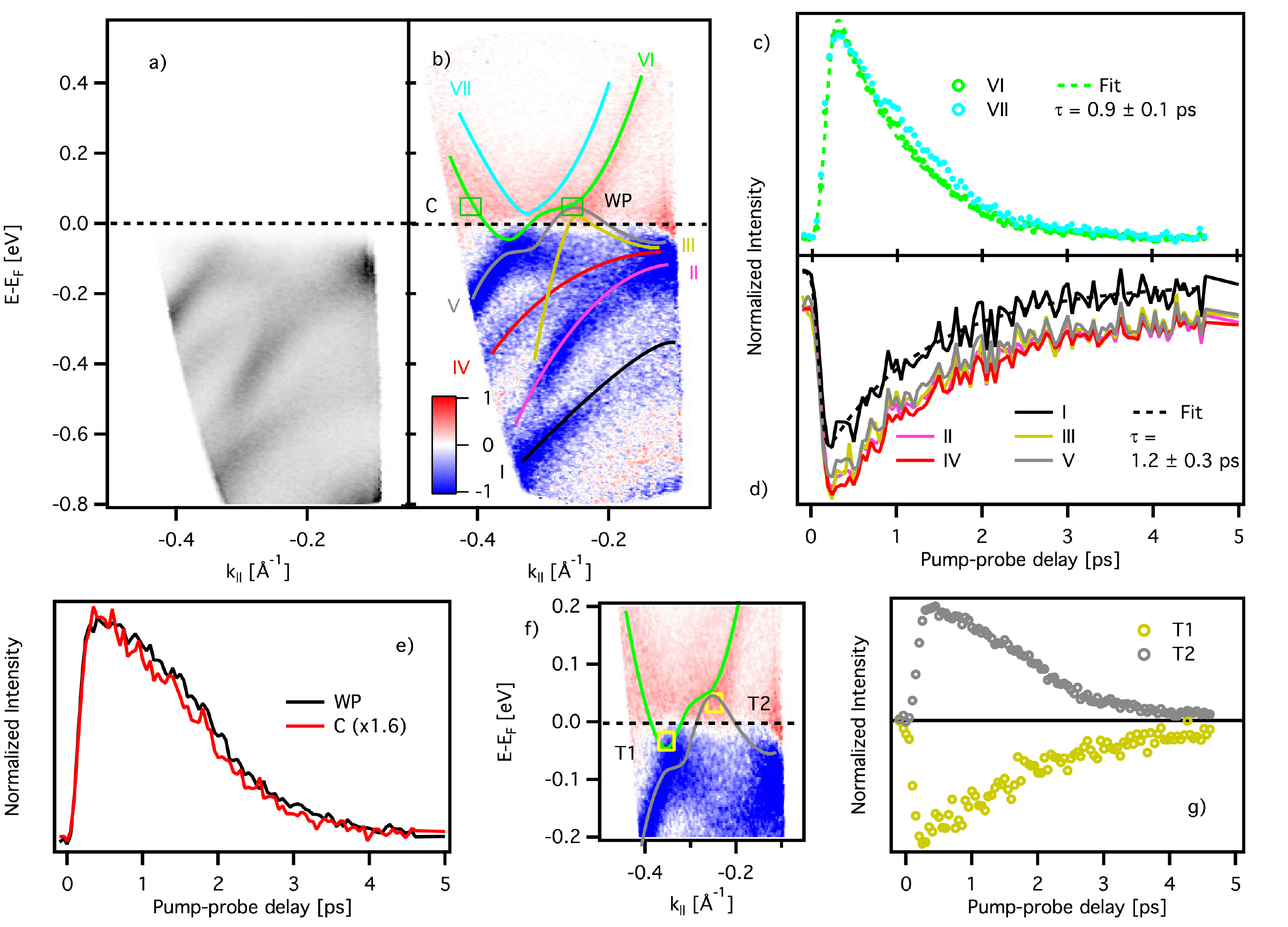}
\caption{\label{dynamic}(Color online) Cut of the band structure along the $\Gamma$X direction (panel a) together with the image difference ($\Delta t=250$ fs - $\Delta t<0$). Lines are guides to the eye showing the integration regions for the bands I-VII, whose normalized temporal evolution of carrier populations is reported in panels c and d. Panel e compared the decay dynamic of the region in the square containing presumably one of the Weyl points (WP), with a control region (C). Panel g compares the electrons and holes dynamic in the electrons and hole pockets T1 and T2 as defined in panel f.}
\end{figure*}

\section{Results and Discussion}
Figure \ref{fermi}(a) shows the Fermi surface as acquired with our system.
With the help of the scheme in figure \ref{fermi}(b) the electron pocket and hole in the $\Gamma$X direction already reported by previous studies are easily recognizable.

Figures \ref{fermi}(c)-\ref{fermi}(e) show cuts of constant momentum from pump-probe ARPES spectra probed with $s$-polarized UV light pulses at 350 fs after the arrival of the pump pulse, \textit{i.e.} $\Delta t=350$ fs.
Cuts of the band structure along the high symmetry direction $\Gamma$X (see Fig.~\ref{fermi}(c)) and at $k_{\text{y}}$=0.05 \AA\textsuperscript{-1} (Fig.~\ref{fermi}(d)) show clearly a broad electron-like band surrounded by a second band that creates the electron pocket below the Fermi level.
Their dispersion moving away from the high symmetry direction, and the dispersion in the perpendicular direction reveal their paraboloid-like shape.
At energies around 0.1-0.2 eV, the two bands expand over pivoting on two distinct points located at [$k_{\text{x}}$=0.23 \AA\textsuperscript{-1}, $k_{\text{y}}$=$\pm$0.013 \AA\textsuperscript{-1}] (Supporting Information - video).
It is also worth noting that the band dispersion in Fig.~\ref{fermi}(e) is quite similar to the one in Fig.~\ref{fermi}(g).
One might ask of whether we have a signature of Weyl points. From what we can tell, the overall band dispersion observed along these points and the fact that the band crossing happens well above the calculated position of the Weyl point suggests that this could not be a spectroscopic evidence of Weyl semimetal. 
Whatsoever, at this point, our experimental resolution does not permit us making conclusive remarks on the topological nature of the observed surface states.
Nevertheless, the analysis of the system in the time domain provides interesting complementary information, in particular on the relaxation of photoexcited carriers.


Figure \ref{dynamic} shows the decay dynamics of photoexcited bands in WTe$_2$ acquired with $p$-polarized pulses along $\Gamma$X.
Bands are labelled following the image shown in Fig.~\ref{dynamic}(b), \textit{i.e.} the difference ARPES image acquired at $\Delta t=250$ fs and an image at equilibrium, right before photoexcitation.
Blue areas correspond to an electronic population depleted upon photoexcitation, while red areas correspond to an excess of electronic population.
Our results are in fairly good agreement with previousely reported calculations and measured band structures\cite{Das2016,Pletikosic2014,Bruno2016}.

In order to follow the decay dynamics, seven regions are delineated as follows: five (I-V) below the Fermi level in correspondence of the occupied bands, and two above the Fermi level in correspondence of the two electron pockets (VI-VII).
Figures \ref{dynamic}(c) and \ref{dynamic}(d) show the intensity integration of the areas surrounding the bands versus pump-probe delay (in supporting information the exact integration area used to extract the band population).
At first glance, we can notice that both electron and hole populations decay exponentially having similar decay times. Single exponential fits to bands VI and III give decay times of 0.9$\pm$0.1 ps and 1.2$\pm$0.3 ps, respectively (Fig.~\ref{dynamic}(b). These values are in perfect agreement with previous time-resolved reflectivity experiments performed by Dai \textit{et~al.} on WTe$_2$\cite{Dai2015}.

Next, we compare the temporal evolution of the excess electronic population in two isoenergy regions around $\sim$30 meV above $E_\text{F}$ marked as WP and C in Fig.~\ref{dynamic}(d).
While both regions belong to the same electron pocket, WP region should be in overall with the Weyl point according to previous discussions.
From our measurements, one can notice a rapid intensity increase at WP shortly after photoexcitation reaching its maximum at $\Delta t\sim$350 fs.
We attribute this behavior to the impact ionization.
Impact ionization in metals and semimetals is known to be the dominant scattering process for carrier relaxation leading to a multiplication of excess electrons at the vicinity of the Fermi level \cite{Fann}.
The process is $k$ and band selective as it acts to minimize the total energy for any given charge carrier.
In presence of a Dirac singularity, the extremely small density of states available for the carriers to cool down establishes a precursor of a population inversion, also known as "bottleneck effect", accumulating electrons and slowing down the charge recombination \cite{Gierz2013,Johannsen2013,Hajlaoui2012}.
Nevertheless, the similar decay rates observed in WP and C suggest that parallel decay channels such as electron-phonon scattering and Auger decay provide adequate energy dissipation to reestablish very quickly a Fermi-Dirac distribution.

Finally, we turn our attention to the comparison of the temporal evolution of the excess carrier densities in the vicinity of the Fermi level.
Figure \ref{dynamic}(g) shows the relaxation dynamics of the electron and hole pockets delimited with the squares T1 and T2 in Fig. \ref{dynamic}(f).
Both populations decay with comparable relaxation times.
A plausible scenario for this behaviour can be the symmetry between the phase space available in proximity of the Fermi level for electrons and holes to relax back to equilibrium.
This means that, even when driven out of equilibrium, the system maintains a remarkable balanced behaviour between electron and hole carriers, strongly supporting the idea that persisting charge compensation is the main mechanism driving the extremely high magnetoresistence in WTe$_2$.

\section{Conclusions}
To summarize, in this paper we showed the unoccupied band structure of WTe$_2$, observing the dispersion of the electron and hole pockets above the Fermi level. Comparing the electron and hole dynamics in the proximity of the Fermi level we found a remarkable balance between the two, lending support to the interpretations invoking charge compensation as a key point to explain the exceptional magnetotransport properties of WTe$_2$. Due also to our experimental resolution, our results could not provide any signature of the presence of a Weyl point, neither from the band structure nor from the relaxation dynamics in the time domain. More studies are required to give a clear conclusion on the topology of the band structure of WTe$_2$.

\begin{acknowledgments}
The FemtoARPES activities were funded by the RTRA Triangle de la Physique, the Ecole Polytechnique, the EU/FP7 under the contract Go Fast (Grant No. 280555). M.C. M.M. and E.P. work was supported by ``Investissement d'avenir Labex Palm'' (Grant No. ANR-10-LABX-0039-PALM) and by the ANR ``Iridoti'' (Grant ANR-13-IS04-0001). The crystal growth work at Princeton University was supported by the NSF MRSEC program grant DMR-1420541
\end{acknowledgments}

\bibliography{apssamp.bib}

\end{document}